\documentclass[twocolumn,showpacs,amsmath,amssymb,prl]{revtex4}

\usepackage{graphicx,float}
\usepackage{subfigure}
\usepackage{dcolumn}
\usepackage{bm}
\usepackage{mathrsfs}

\def\cosp{\mathcal{C}_p}

\def\sinp{\mathcal{S}_p}

\allowdisplaybreaks[2]

\begin{document}

\title{Ground state cooling of a nanomechanical resonator in the non-resolved regime via quantum interference}

\author{Keyu Xia}

\author{J\"{o}rg Evers}%

\affiliation{Max-Planck-Institute f\"{u}r Kernphysik, Saupfercheckweg 1, D-69117 Heidelberg, Germany}

\date{\today}

\begin{abstract}
Ground state cooling of a nanomechanical resonator coupled to a superconducting flux qubit is discussed. By inducing quantum interference to cancel unwanted heating excitations, ground state cooling becomes possible in the non-resolved regime. The qubit is modelled as a three-level system in $\Lambda$ configuration, and the driving fluxes are applied such that the qubit absorption spectrum exhibits electromagnetically induced transparency, thereby cancelling the unwanted excitations. As our scheme allows to apply strong cooling fields, fast and efficient cooling can be achieved. 
\end{abstract}



\maketitle

Nanomechanical resonators (NAMR) currently attract considerable interest because of their combination of high natural frequencies and large quality factors together with a wide range of potential applications~\cite{review}. Among them are measurements of displacement at the quantum limit~\cite{qlimit}, mass measurements~\cite{NanoLett5p925}, biological sensing~\cite{NatureNanotech3p501} and the observation of quantum mechanical phenomena in mesoscopic objects~\cite{review}.
To fully utilize the properties of NAMRs or to observe mesoscopic quantum phenomena, it is typically necessary to cool the NAMR to the mechanical ground state. Thus it is not surprising that a number of different approaches for cooling micro- and nanomechanical resonators have been proposed theoretically~\cite{cooper,PRL92p075507,qtheory,PRL100p047001,PRB68p235328,PRB76p014511} and also demonstrated experimentally~\cite{PRL101p197203,Nature444p75,Nature443p193,PRL99p017201,PRL101p033601,NaturePhys4p415,Nature444p67}.
For micromechanical resonators, cavity-assisted radiation pressure cooling has been intensely studied~\cite{qtheory,NaturePhys4p415,review,Nature444p67,n1}. A different approach is active feedback cooling, which however typically requires difficult and precise measurements in real time of the displacement of the resonator~\cite{PRB68p235328,Nature444p75,PRL99p017201,PRL101p033601}. Cavity-based schemes are limited by diffraction, if the size of the resonator is small compared to the wavelength of the light. For NAMR, it has been proposed to achieve cooling by periodic coupling to a superconducting qubit (SQ) such as a Cooper pair box (CPB)~\cite{cooper} or to a three-level flux qubit~\cite{PRL100p047001}. Both techniques rely on a strong resonant interaction between resonators and the qubit.
Recently, sideband cooling of micro- and nanomechanical resonators has attracted considerable interest. For example, cooling a NAMR has been proposed by embedding a quantum dot in the resonator~\cite{PRL92p075507}, and it was observed in a microresonator~\cite{NaturePhys4p415} and in a transmission line resonator~\cite{PRL101p197203,PRB76p014511}. Also, a quantum theory of cooling has been developed~\cite{qtheory}.

A number of problems associated with cooling NAMR are shared by laser cooling of atoms or ions. In particular, ground state sideband cooling is possible only in the resolved regime \cite{qtheory}, in which the motional sidebands are resolved from the linewidth of the involved transitions~\cite{RMP75p281,qtheory}. This has been realized recently in few systems~\cite{PRL101p197203,n1}, but still this regime typically is difficult to achieve, and limits the accessible parameter range. To overcome this limit in atomic systems, a cooling scheme based on electromagnetically induced transparency (EIT)~\cite{RMP77p633} has been proposed~\cite{PRL85p4458,EPL68p370} and experimentally verified in ions~\cite{APB73p807}. EIT cooling works in the non-resolved regime, but suppresses the carrier excitation without change in the motional quantum number. This is achieved by designing the optical properties of the target system in such a way that absorption vanishes at the carrier transition frequency.

In this Letter, we discuss ground state cooling of a NAMR in the non-resolved regime. The NAMR is embedded in the loop of a flux qubit. The qubit is modelled as a three-level quantum system in $\Lambda$ configuration, and time-dependent magnetic fluxes (TDMF) are applied to the qubit in such a way that detrimental carrier excitations  without change in the motional quantum number  ($|n\rangle\rightarrow |n\rangle$) are suppressed by quantum interference.
We find that the cooling limit of the NAMR has two contributions. One originates from the scattering of the cooling fields, whereas the other one arises from the equilibrium phonon number of the environmental thermal bath. Our interference-based cooling scheme extends to strong cooling fields, and thus enables rapid cooling to a high occupancy of the mechanical ground state.
Unlike backaction cooling~\cite{qtheory}, no significant coherent shift occurs in the final occupancy of resonator in EIT cooling. Our system allows for  large Lamb-Dicke (LD) parameters via controlling the applied magnetic field or the working point, which leads to rapid cooling to a high occupancy of the motional ground state of the resonator. An experimental implementation is facilitated by a rather small required input power.

\begin{figure}[t]
\centering
\includegraphics[width=\columnwidth]{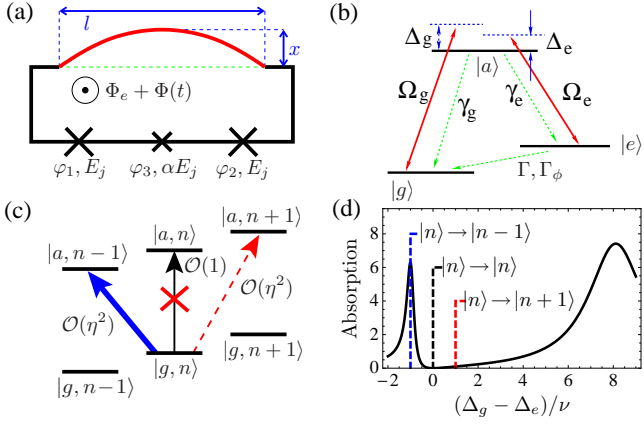}
\caption{\label{fig:system}(Color online) (a) A nanomechanical resonator (red curve) coupling to a superconducting flux qubit. (b) The level diagram of the flux qubit. (c) Effective level scheme in the Lamb-Dicke (LD) limit. (d) Example absorption spectrum of the cooling field. Indicated are the transition frequencies for the different absorption channels in the LD limit.}
\end{figure}

%
We start with a discussion of the main idea. As shown in Fig.~\ref{fig:system}(a), the NAMR is part of the qubit loop. As it vibrates, the area of the flux qubit changes, which leads to a coupling of NAMR and qubit. The qubit is modelled as a three-level system as shown in Fig.~\ref{fig:system}(b). The driving field $\Omega_g$ acts as cooling field, while the field $\Omega_e$ is a control field. The relevant qubit-resonator level scheme in the LD limit is shown in Fig.~\ref{fig:system}(c). So-called carrier transitions $|g,n\rangle \to |a,n\rangle$ do not change the resonator state $|n\rangle$, while sideband transitions $|g,n\rangle \to |a,n\pm 1\rangle$ do. Both carrier transitions and sideband transitions $|g,n\rangle \to |a,n+ 1\rangle$ with subsequent spontaneous decay on average induce heating.
In resolved sideband cooling, in which  the resonator phonon energy $\hbar \nu$  exceeds the transition width $\hbar\gamma$, a low cooling limit is achieved since the cooling sideband can almost selectively be driven. In the non-resolved regime ($\nu < \gamma$), scattering on the carrier transition dominates the heating processes, prohibiting ground state cooling. The relative probabilities of the excitation channels are determined by the absorption spectrum on the cooling transition, which for a two-level system is of Lorentzian shape.
The principle idea of EIT cooling is to modify this absorption spectrum such that the carrier transitions are eliminated. For EIT, an additional  driving field is applied as shown in Fig.~\ref{fig:system}(b)~\cite{RMP77p633}.
If both fields have equal detuning from the respective qubit transition frequencies, it turns out that the qubit is driven into a coherent superposition of the two ground states $|g\rangle$ and $|e\rangle$. From this superposition, excitation to $|a\rangle$ is possible either via $|e\rangle$ or $|g\rangle$. These channels, however, destructively interfere, such that no absorption occurs even though both fields are applied. This phenomenon is known as  EIT~\cite{RMP77p633}. For suitable field parameters, a cooling field absorption spectrum as shown in Fig.~\ref{fig:system}(d) can be achieved. Carrier excitations are suppressed since no cooling field absorption occurs at the corresponding frequency due to EIT. This way, ground state cooling is achieved even in the non-resolved regime.

We now proceed with a quantitative analysis of the qubit-NAMR system. We assume that the NAMR vibrates in its fundamental mode and in the plane of the loop. It has an effective mass $M_{eff}$, length $l$, frequency $\nu$, and amplitude of the fundamental oscillation mode $x$. The quantum NAMR has Hamiltonian $H_r=\hbar\nu b^\dag b$, annihilation operator $b$, and zero point fluctuation $X_0=\sqrt{\hbar/2M\nu}$. 
%
The flux qubit consists of a superconducting loop with three Josephson junctions. Two junctions are identical, while the third is smaller by a factor of $\alpha$. The two larger junctions have equal Josephson energies $E_{J1}=E_{J2}=E_J$ and capacitances $C_{J1}=C_{J2}=C_J$, while for the third one $E_{J3}=\alpha E_J$ and $C_{J3}=\alpha C_J$, with $\alpha <1$. The gauge-invariant phase drops across the three junctions in the qubit loop are $\varphi_1$, $\varphi_2$ and $\varphi_3$. The whole structure is exposed both to a constant magnetic field $\textbf{B}$ perpendicular to the plane and to driving microwave fields giving rise to time dependent magnetic fluxs (TDMFs) $\Phi(t)$. Introducing $\varphi_p=(\varphi_1+\varphi_2)/2$ and $\varphi_m=(\varphi_1-\varphi_2)/2$ as the coordinates,
the Hamiltonian can be written as $H=H_0+H_r+H_I$, with $H_0$ and $H_r$ as the free Hamiltonian of the qubit and the NAMR, and $H_I$ as the interaction part. The last term in $H_I$ is crucial to EIT cooling because it includes an interaction of resonator, qubit and field. We define $\cosp=\cos(2\varphi_p+2\pi f)$ and $\sinp=\sin(2\varphi_p+2\pi f)$, and obtain
$H_0=P_m^2/(2M_m)+P_p^2/(2M_p)-2E_J \cos\varphi_m \cos\varphi_p-\alpha E_J \cosp$
as qubit part, and 
$H_I=\alpha/(1+2\alpha)\bar{\Phi}[Bl\dot x+\dot \Phi (t)]P_p
+\alpha E_J \bar{\Phi}[Blx+\Phi(t)] \sinp +\alpha E_J\bar{\Phi}^2[Blx+\Phi (t)]^2 \cosp/2$
as the interaction part. Here, we have introduced the momenta $P_m=-i\hbar\partial/\partial \varphi_m$ and $P_p=-i\hbar\partial/\partial\varphi_p$, and the effective masses of qubit  $M_m=2C\bar{\Phi}^2$ and $M_p=(1+2\alpha)M_m$. The bias  $f=\Phi_e/\Phi_0$ where $\bar{\Phi}=2\pi/\Phi_0$ with $\Phi_0$ as the flux quantum, and $\Phi_e$ is the static bias flux corresponding to the equilibrium position $x=0$ of the NAMR. We further assumed  moderate TDMF  such that $\xi = \bar{\Phi}(Blx+\Phi (t))$ is small, and expanded corresponding trigonometric functions to second order in $\xi$.
As indicated in Fig.~\ref{fig:system}(b), we apply two TDMF with different frequencies $\omega_{Lg}$, $\omega_{Le}$ and amplitudes $A_g$, $A_e$.
The corresponding Rabi frequencies are
$\hbar\Omega_j=\alpha E_J\bar{\Phi}A_j\langle a|\sinp |j\rangle/2$, 
where we have dropped a small contribution from the momenta.
The LD parameters are defined as $\eta_{LD}=|\eta_g-\eta_e|$, where
$\eta_j=BlX_0\bar{\Phi}\langle a|\cosp|j\rangle/\langle a|\sinp|j\rangle$.

Applying the rotating-wave, the Born-Markov and the LD approximation ($\eta_{LD}\ll 1$), we obtain 
\begin{subequations}
\begin{align}
\label{eq:MEq}
&\dot \rho =-\frac{i}{\hbar}[\tilde{H}_{0}+\tilde{H}_I,\rho]\nonumber\\
&+ \mathscr{L} \left(\gamma_g, |g\rangle \langle a| B_g\right)\rho+\mathscr{L} \left(\gamma_e,  |e\rangle \langle a| B_e\right)\rho \nonumber\\
&+\mathscr{L} \left(\Gamma,  |g\rangle \langle e| B_3\right)\rho+\mathscr{L} \left(\Gamma_\phi/2,|e\rangle \langle e|-|g\rangle \langle g|\right)\rho \nonumber\\
&+[N(\nu)+1]\mathscr{L} \left(\nu/Q,b\right)+N(\nu)\mathscr{L} \left(\nu/Q,b^\dag\right)\,,\\
&\tilde {H}_0 =-\hbar \Delta_g |g\rangle\langle g|-\hbar \Delta_e |e\rangle \langle e|+\hbar \nu b^\dag b¸\,, \\
&\tilde {H}_I =\hbar \Omega_g B_g |a\rangle\langle g|+\hbar \Omega_e B_e |a\rangle\langle e|+H.c. \,,\\
&\mathscr{L} \left(\tilde{\gamma},A \right)\rho =\tilde{\gamma}/2\left\{2A\rho A^\dag-A^\dag A \rho -\rho A^\dag A \right\} \,,
\end{align}
\end{subequations}
with $B_j=\mathbb{I}+\eta_j (b+b^\dag)$ for $j\in\{g,e,3\}$. Initially, the NAMR occupation is $N_i = N(\nu)=\left[\exp(\hbar\nu/k_B T)-1\right]^{-1}$ due to the thermal environment of temperature $T$.  The detunings between the TDMF and the corresponding transition frequencies are $\Delta_g=\omega_{ag}-\omega_{Lg}$ and $\Delta_e=\omega_{ae}-\omega_{Le}$. We have redefined the transition frequencies to include negligible level shifts.
The decay rates are defined as in Fig.~\ref{fig:system}(b), and the pure dephasing rate of transition $|g\rangle \leftrightarrow |e\rangle$ is denoted by $\Gamma_\phi$. 
In Eq.~(\ref{eq:MEq}) the second, third and forth terms describe the spontaneous emission of the flux qubit.
The fifth term considers an additional pure dephasing.
The final terms include a thermal bath, since NAMR operate at frequencies with non-negligible thermal mode excitation. $Q$ is the NAMR quality factor.

In the LD limit, the qubit degrees of freedom can be adiabatically eliminated to derive a rate equation for the vibrational NAMR states. For the analytical analysis, we neglect the decay from the excited state $|e\rangle$ to the ground state $|g\rangle$ and their decoherence because the rates $\Gamma$ and $\Gamma_\phi$ can be designed much smaller than the rates $\gamma_g$ and $\gamma_e$. In two-photon resonance $\Delta_g=\Delta_e=\Delta$, the rate equation for the average number of phonons $\langle n\rangle=\sum_{n=0}^{\infty}n\langle n|\rho|n\rangle$ of the vibrational number states $|n\rangle$ is
$d/dt \: \langle n\rangle=-\left(W +\nu/Q\right)\langle n\rangle 
+A_+ + \nu N(\nu)/Q+\delta A_+$.
Here, $\delta A_+=(\eta_g^2 \gamma_g+\eta_e^2\gamma_e)\rho_a^{(ss)}/2+\eta_3^2 \Gamma \rho_{e}^{(ss)}/2$ is negligible since $\rho_a^{(ss)}$, $\Gamma$ and $\eta_3$ are small. $\rho_a^{(ss)}$ and $\rho_e^{(ss)}$ are the steady-state population of auxiliary state $|a\rangle$ and $|e\rangle$ in the absence of the NAMR. We identify $W = A_- - A_+$ with a net cooling rate in the zero-temperature case~\cite{PRL92p075507}, such that cooling requires $W+\nu/Q>0$.  The two transition rates $A_\pm$ describing the heating and cooling excitations are given by [$\Omega=(\Omega_g^2+\Omega_e^2)^{1/2}$]
\begin{equation}
A_\pm=\frac{4\eta_{LD}^2 \Omega_g^2 \Omega_e^2}{\Omega^2 \gamma}\frac{\gamma^2 \nu^2}{\gamma^2 \nu^2+4\left[\Omega^2-\nu (\nu\pm\Delta)\right]^2}\,.
\end{equation}
The steady state evaluates to $n_{ss}=\nu N_i/QW+A_+/W$ and is minimal for $\Omega=\sqrt{\nu (\nu-\Delta)}$ and $\Delta<0$. Then, the cooling rate scales as $W_{max}\sim 4\eta_{LD}^2 \Omega_e^2\Omega_g^2/(\Omega^2 \gamma)$.
To analyze the cooling dynamics further, we assume $|\Delta| > \nu$, define  $r = \Omega_e/\Omega_g$, and distinguish the two cases of
weak ($r\gg 1$) and strong cooling fields ($r\approx1$). We find
\begin{subequations}\label{eq:nss}
\begin{align}
n_{ss}(r\gg1)&\approx\frac{\gamma r^2N_i}{4\eta_{LD}^2 Q|\Delta|}+\frac{\gamma^2}{(4\Delta)^2}\,,\\
n_{ss}(r\approx1)&\approx \frac{\gamma N_i}{\eta_{LD}^2 Q|\Delta|}+\frac{\gamma^2}{(4\Delta)^2}\,.\label{strong}
\end{align}
\end{subequations}
Similar to sideband cooling,  $n_{ss}$ has two contributions. The first arises from the environment and is proportional to $N_i$, while the second one is due to cooling field scattering.
But there are crucial differences to other cooling approaches. Even though we operate in the non-resolved regime, the field scattering contribution is inversely proportional $\Delta^2$, allowing for ground state cooling. In contrast, this is only possible in resolved sideband cooling~\cite{PRL92p075507,qtheory}.
Also, other than in backaction cooling~\cite{qtheory}, no significant coherent shift of the final phonon number occurs in EIT cooling. Finally,  in EIT cooling, the environmental contribution to the cooling limit is suppressed by a factor of $\Delta$ relative to sideband cooling \cite{PRL92p075507} and backaction cooling. For the latter comparison, we define an effective LD parameter $\eta_{LD}=\eta\sqrt{n_{max}}$, where $\eta=(x_0/\nu)\, d/dx\, \omega_c$ with resonance frequency $\omega_c$ of the cavity and $n_{max}$ as the photon number in resonance~\cite{qtheory}.

We now turn to our results based on our analytical analysis and on a numerical solution of the full system Eq.~(\ref{eq:MEq})~\cite{JOB1p424}.
We choose the flux qubit parameters  based on the experimental work~\cite{PRB76p214503} as  $\alpha=0.7$, $E_J=200$GHz, and $E_J/E_C$=50, where $E_C$ is the charging energy of the junction. We assume a bias flux $f=0.5005$, and calculate transition frequencies $\omega_{eg}\approx 2\pi\times 4.89$GHz, $\omega_{ag}\approx2\pi\times 30.68$GHz and $\omega_{ae}\approx 2\pi\times25.79$GHz.
%
The decay and pure dephasing rates are taken as the measured values $\Gamma\sim 2\pi\times 2$MHz and $\Gamma_\phi\sim 2\pi\times 4$MHz at $f=0.5005$~\cite{PRB76p214503}. The linewidth of  state $|a \rangle$ is given by $\gamma=\gamma_g +\gamma_e$. Since $\gamma$ was not measured in~\cite{PRB76p214503}, we assume $\gamma=50\Gamma$~\cite{PRL93p087003,QIP8p261}. As the final phonon number is insensitive to $\gamma_g/\gamma_e$, we choose $\gamma_g=\gamma_e=\gamma/2$.
The LD parameters are $\eta_g=28.19BlX_0\bar{\Phi}$ and $\eta_e=-0.10BlX_0\bar{\Phi}$ and $\eta_3=BlX_0\bar{\Phi}\langle e|\cosp|g\rangle/\langle e|\sinp|g\rangle=0.02BlX_0\bar{\Phi}$, respectively.
Rabi frequencies $\Omega_e\sim \gamma$ can be achieved with input power  $\leq 0.5~\mu W$~\cite{PRB76p214503}. This power can be further reduced via a larger mutual inductance between qubit and driving circuit.
The NAMR is taken as a double clamped resonator with size $25~\mu$m$\times 100$nm$\times 120$nm, effective mass $M_{eff}\lesssim 2$pg, resonance frequency $\nu=2\pi\times 25$MHz  and quality factor $Q=5\times 10^4$~ \cite{arXiv08034007,Nature443p193,NanoLett7p1728}. We assume a magnetic field $B\lesssim 3$T, which is resonable in NbN-based qubits, to obtain $\eta_{LD}=0.0566$.
%

\begin{figure}[t]
\centering
\includegraphics[width=\columnwidth]{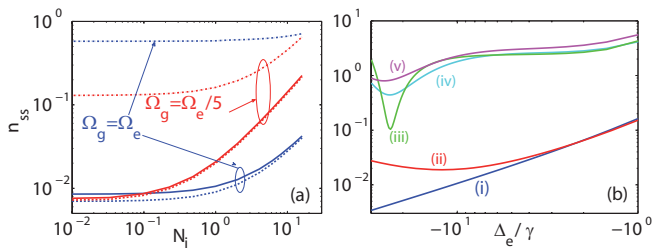}
\caption{\label{fig2}(Color online) 
(a) Cooling limit $n_{ss}$ against initial phonon number $N_i$. Dashed lines show Eq.~(\ref{eq:nss}), and solid lines show full numerical results with $\Gamma=0$, $\Gamma_\phi=0$. The dash-dotted lines show numerical results including decoherence $\Gamma=0.02\gamma$, $\Gamma_\phi=0.04\gamma$ as measured in~\cite{PRB76p214503}.
Results are shown for two different cooling field strenghts. Parameters are $\Omega_e = \sqrt{\nu(\nu-\Delta)}$ [$\sqrt{\nu(\nu-\Delta)/2}$] for weak [strong] cooling fields and $\Delta=-3\gamma$, except for the dash-dotted curve with $\Omega_e = \Omega_g$ which shows $\Delta_g=-2.85\gamma$, $\Delta_e=-3\gamma$, and $\Omega_e = 0.53\gamma$.
(b) Cooling limit $n_{ss}$ as a function of the detuning $\Delta_e$. Parameters are $N_i=16$ and $\Omega_e = \Omega_g$. Shown are (i) Eq.~(\ref{eq:nss}), and (ii) numerical results  for $\Gamma=\Gamma_\phi=0$, both with $\Omega_e = \Omega_e^{opt}:=\sqrt{\nu(\nu-\Delta_e)/2}$ and $\Delta_e = \Delta_g$. The other curves are numerical results for $\Delta_g=0.99\Delta_e$, $\Gamma=0.02\gamma$, and (iii) $\Gamma_\phi=0$, $\Omega_e=0.58 \Omega_e^{opt}$, (iv) $\Gamma_\phi=0.04\gamma$,  $\Omega_e=0.64\Omega_e^{opt}$ and (v) $\Gamma_\phi=0.08\gamma$,  $\Omega_e=0.64 \Omega_e^{opt}$.%
}
\end{figure}

First, we study the dependence of the cooling limit $n_{ss}$ on the temperature of the environmental bath. Corresponding results are shown in Fig.~\ref{fig2}(a) for both weak and strong cooling fields.
In the ``small $N_i$-large $\eta_{LD}^2 Q$'' regime, the dominant contribution to the cooling limit arises from the scattering of the cooling TDMF. Neglecting the environmental contribution, the steady-state phonon number $n_{ss}$ then simplifies to $n_{ss}=\gamma^2/(4\Delta)^2$. This result is independent of $r$, unless $r$ becomes large enough to outweigh the small ratio $N_i/(\eta_{LD}^2 Q)$.
Consequently, $n_{ss}$ saturates to a constant value towards smaller $N_i$. 
In the second ``large $N_i$-small $\eta_{LD}^2 Q$'' regime, the contribution proportional to $N_i$ arising from the environment dominates the cooling limit. 
In Fig.~\ref{fig2}(a), we see that our approximate analytical results agree well with the numerical results for negligible decoherence on the ground state transition. 
The dash-dotted curves in Fig.~\ref{fig2}(a) show the numerical results with decay and dephasing rates as measured in~\cite{PRB76p214503}. With decoherence, field parameters slightly modified from the analytical predictions for the decoherence-free case improve the cooling. For example, assuming $N_i=16$ corresponding to a temperature $T=20$mK, a weak [strong] cooling field cools the system down to $n_{ss}=0.65$ [$n_{ss}=0.71$] in this case, i.e., close to the motional ground state.

In Fig.~\ref{fig2}(b), the dependence of the steady state phonon number $n_{ss}$ on the detuning $\Delta$ is shown for  $T =20$mK ($N_i=16$). From our analytical results, we expect  stronger cooling fields to be particularly effective if larger detunings are used, since the environmental contribution is suppressed by the detuning $\Delta$, see Eq.~(\ref{strong}).
Therefore, we focus on stronger cooling fields ($r=1$).  As before, good agreement between theory and numerical results is obtained for the case with negligible ground state decoherence. As predicted from the analytical results, the cooling limit decreases with increasing detuning $|\Delta|$.  Note, however, that our analytical results become invalid for $|\Delta|\gtrsim 10\gamma$ because the system then cannot be excited efficiently.  For the numerical results including decoherence, we obtimize the chosen driving field parameters slightly away from the parameters suggested from the analytical calculation. For the decoherence parameters measured in~\cite{PRB76p214503} shown in curve (iv), a cooling limit of $n_{ss}=0.44$ is achieved, corresponding to a ground state occupancy of $70\%$. As expected, higher [lower] decoherence rates lead to higher [lower] final phonon number. 
Note that in principle, $\Gamma$ and $\Gamma_\phi$ can be smaller by several orders of magnitude than $\gamma$ in our system~\cite{PRL100p047001,QIP8p261}, leading to a cooling performance as in curve (ii) of Fig.~\ref{fig2}.

Finally, we note that our scheme can simultaneously cool several resonator modes~\cite{APB73p807}. For example, assuming  $\Gamma=0.02\gamma$, $\Gamma_\phi=0.04\gamma$, $\Delta_e = -3\gamma$, $\Delta_g = -2.86\gamma$, and $\Omega_g=\Omega_e=0.694\gamma$, the point of vanishing absorption moves between  two modes $\nu_1=\gamma/4$ and $\nu_2=\gamma/2$. We find that one can then simultaneously cool both modes from $N_i=16$ to about $n_{ss}=1$, i.e., $50\%$ ground state occupation. As in the single-mode case, cooling improves further with decreasing $\Gamma$ and $\Gamma_\phi$.

In summary, we studied ground state cooling of a nanomechanical resonator coupled to a flux qubit in the non-resolved regime. Efficient cooling is achieved because detrimental carrier excitations are suppressed by quantum interference.


\end{document}